\documentclass[12pt,a4paper,oneside,onecolumn,bibliography=totocnumbered]{article}  
\pdfoutput=1 
\usepackage{ucs}
\usepackage{cite}
\usepackage{amsmath} 
\usepackage{comment}
\usepackage[utf8x]{inputenc}
\newcommand\rf[1]{(\ref{eq:#1})}
\newcommand\lab[1]{\label{eq:#1}}

\newcommand\br{\begin{eqnarray}}
\newcommand\er{\end{eqnarray}}
\newcommand\be{\begin{equation}}
\newcommand\ee{\end{equation}}

\newcommand\bc{\begin{center}}
\newcommand\ec{\end{center}}

\numberwithin{equation}{section}

\def\beq{\begin{equation}}
\def\eeq{\end{equation}}
\def\ber{\begin{eqnarray}}
\def\eer{\end{eqnarray}}

\def \lleq {\lower0.9ex\hbox{ $\buildrel < \over \sim$} ~}
\def \ggeq {\lower0.9ex\hbox{ $\buildrel > \over \sim$} ~}
\def\be{\begin{equation}}
\def\ee{\end{equation}}
\def\ba{\begin{eqnarray}}
\def\ea{\end{eqnarray}}
\usepackage{graphicx} 
\usepackage{color} 
\usepackage{subfigure} 
\usepackage{float} 
\usepackage{capt-of} 
\usepackage{sidecap} 
\sidecaptionvpos{figure}{c} 
\usepackage{caption} 
\usepackage{commath} 
\usepackage{graphicx, amsmath, amsthm, latexsym, amssymb, amsfonts, epsfig, float, enumerate, color, listings,  graphicx, fancyhdr}
\usepackage{fouriernc}
\usepackage[T1]{fontenc}
\usepackage{cancel} 
\usepackage[table,xcdraw]{xcolor} 
\usepackage{anysize} 					
\usepackage[final]{pdfpages}

\marginsize{2cm}{2cm}{2cm}{2cm} 

\usepackage{appendix}

\usepackage[colorlinks=true,plainpages=true,citecolor=blue,linkcolor=blue]{hyperref}
\usepackage{fancyhdr} 
\usepackage{listings} 
\definecolor{dkgreen}{rgb}{0,0.6,0} 
\definecolor{gray}{rgb}{0.5,0.5,0.5} 
\newcommand{\sen}{\operatorname{\sen}}

\begin{document}

\begin{center}
\vspace*{2mm}
\rule[0.5ex]{\linewidth}{2pt}\vspace*{-\baselineskip}\vspace*{3.2pt}
\rule[0.5ex]{\linewidth}{1pt}\\
[\baselineskip]{\Huge Unified Dark Energy and Dark Matter from Dynamical Spacetime Cosmology}\\[3mm]
\rule[0.5ex]{\linewidth}{1pt}\vspace*{-\baselineskip}\vspace{3.2pt}
\rule[0.5ex]{\linewidth}{2pt}\\
 [9mm]
{\large \textit{Thesis submitted in partial fulfillment of the requirements for \\ [2mm]
the degree of “DOCTOR OF PHILOSOPHY”
}}\\ [2mm]
\vspace{8mm}
{\large By}\\
\vspace{2.5mm}
{\large\textsc{David Benisty}}\\
\vspace{10mm}
\includegraphics[scale=0.9]{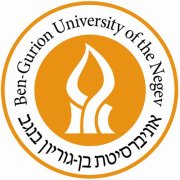}\\
\vspace{9mm}
{\large Submitted to the Senate of Ben-Gurion University
of the Negev
}\\
\vspace{11mm}
{\large\textsc{JANUARY 2021}}\\
\vspace{12mm}
{\large\textsc{Beer Sheva, Israel}}
\end{center}																									
\newpage													
\mbox{} 
\newpage
\begin{center}
\vspace*{3mm}
\rule[0.5ex]{\linewidth}{2pt}\vspace*{-\baselineskip}\vspace*{3.2pt}
\rule[0.5ex]{\linewidth}{1pt}\\
[\baselineskip]{\Huge Unified Dark Energy and Dark Matter from Dynamical Spacetime Cosmology}\\[3mm]
\rule[0.5ex]{\linewidth}{1pt}\vspace*{-\baselineskip}\vspace{3.2pt}
\rule[0.5ex]{\linewidth}{3pt}\\
\vspace*{5mm}
{\large \textit{Thesis submitted in partial fulfillment of the requirements for \\ [2mm]
the degree of “DOCTOR OF PHILOSOPHY”
}}\\ [2mm]
\vspace{6.5mm}
{\large By}\\
\vspace{2.5mm}
{\large\textsc{David Benisty}}\\
\vspace{6.5mm}
{\large This work was carried out under the supervision of}\\
\vspace{2.5mm}
{\large\textsc{Prof. Eduardo Guendelman}}\\
\vspace{8mm}
{\large The Physics Department\\
\textsc{Faculty of Natural Sciences}}\\
\vspace{11mm}
\begin{minipage}{10cm}
{\textit{Approved by the advisor:}} 
\\
{\textit{Approved by the Dean of the \\Kreitman School of Advanced Graduate Studies:\\ ...............................................................}}
\end{minipage}\\
\vspace{11mm}
{\large\textsc{JANUARY 2021}}\\
\vspace{12mm}
{\large\textsc{Beer Sheva, Israel}}
\end{center}						
\newpage													
\mbox{} 
\newpage

\tableofcontents 

\newpage													
\mbox{} 
\newpage
\section{Research Student's Affidavit}
David Benisty, whose signature appears below, hereby declare that:
\begin{itemize}
\item I have written this Thesis by myself, except for the help and guidance offered by my Thesis Advisors.

\item  The scientific materials included in this Thesis are products of my own research, culled from the period during which I was a research student.

\end{itemize}

Date: 20.01.2021

Student's name: David Benisty

Signature: 

\newpage													
\mbox{} 
\newpage
\section{Abstract} 
A model of unified dark matter and dark energy based on a   {Dynamical Spacetime Theory (DST)} is studied.   {By introducing a Dynamical Spacetime vector field $\chi_\mu$, a conservation of an energy momentum tensor $T^{\mu\nu}_{(\chi)}$ emerges.} The action allows for two different potentials, while one represents a dark energy. For constant potentials, the cosmological solution yields a non singular bouncing solutions  {that rapidly approaches} the $\Lambda$CDM model. The Dynamical Time corresponds to the cosmic time as well. The theory fits with the late time expansion data of the  {Universe}.  {With higher dimensions a mechanism for inflation and compactification} appears,    {with exponential growth for some dimensions and exponential contraction of the others.} By demanding that the Dynamical Spacetime vector field be a gradient of a scalar the DST becomes a theory with diffusive interacting, which asymptotically  {returns to the $\Lambda$CDM model} as a stable point. These formulations lead to scenarios which address our understanding about the origin of the  {Universe}.

\newpage													
\mbox{} 
\newpage

\section{List of Papers}
The following papers are included in the thesis:
 \begin{enumerate}
   \item   \textbf{D.~Benisty} and E.~I.~Guendelman,

  ``Unified dark energy and dark matter from dynamical spacetime,'' 
  
  Phys.\ Rev.\ D {\bf 98}, no. 2, 023506 (2018),
  
  arXiv:1802.07981
    \item  F.~K.~Anagnostopoulos, \textbf{D.~Benisty}, S.~Basilakos and E.~I.~Guendelman,

  ``Dark energy and dark matter unification from dynamical spacetime: observational constraints and cosmological implications,''

  JCAP {\bf 1906}, 003 (2019),

  arXiv:1904.05762

\item \textbf{D.~Benisty} and E.~I.~Guendelman,

 ``Inflation compactification from dynamical spacetime,''

  Phys.\ Rev.\ D {\bf 98}, no. 4, 043522 (2018),

  arXiv:1805.09314.
 
  \item     \textbf{D.~Benisty} and E.~I.~Guendelman,

 ``Interacting Diffusive Unified Dark Energy and Dark Matter from Scalar Fields,'' 
  
    Eur.\ Phys.\ J.\ C {\bf 77}, no. 6, 396 (2017),
  
 arXiv:1701.08667.

\item   \textbf{D.~Benisty}, E.~Guendelman and Z.~Haba,

 ``Unification of dark energy and dark matter from diffusive cosmology,''

  Phys.\ Rev.\ D {\bf 99}, no. 12, 123521 (2019),

 arXiv:1812.06151.

\end{enumerate}

\newpage													
The following are additional papers which did not fit within the story line of this thesis and are not included.

 \begin{enumerate}
 \item  \textbf{D.~Benisty}, D.~Vasak, J.~Kirsch and J.~Struckmeier,
  ``Low-redshift constraints on covariant canonical Gauge theory of gravity,'' Eur.\ Phys.\ J.\ C {\bf 81}, no. 2, 125 (2021), arXiv: 2101.07566.

 \item   \textbf{D.~Benisty} and D.~Staicova,
  ``Testing Late Time Cosmic Acceleration with uncorrelated Baryon Acoustic Oscillations dataset,''
  Astron.\ Astrophys.\  {\bf 647}, A38 (2021), arXiv:  2009.10701.
 
   \item   \textbf{D.~Benisty},
  ``Quantifying the $S_8$ tension with the Redshift Space Distortion data set,''
  Phys.\ Dark Univ.\  {\bf 1}, 100766 (2021), arXiv: 2005.03751.
  
  \item   \textbf{D.~Benisty}, E.~I.~Guendelman, A.~Kaganovich, E.~Nissimov and S.~Pacheva,
  ``Non Canonical Volume Form Formulation of Modified Gravity Theories and Cosmology,''
  Eur.\ Phys.\ J.\ Plus {\bf 136}, no. 1, 46 (2021), arXiv: 2006.04063.

\item   \textbf{D.~Benisty} and E.~I.~Guendelman,
  ``The Local Group as a test system for Modified Newtonian Dynamics,'' Phys.\ Dark Univ.\  {\bf 30}, 100708 (2020), arXiv: 2007.13006.
\item   \textbf{D.~Benisty} and E.~I.~Guendelman,
  ``Quintessential Inflation from Lorentzian Slow Roll,''
  Eur.\ Phys.\ J.\ C {\bf 80}, no. 6, 577 (2020),
  arXiv: 2006.04129.
 
\item  \textbf{D.~Benisty} and E.~I.~Guendelman,
  ``Lorentzian Quintessential Inflation,''
  Int.\ J.\ Mod.\ Phys.\ D {\bf 29}, no. 14, 2042002 (2020), arXiv: 2004.00339.
 
   \item   \textbf{D.~Benisty}, E.~I.~Guendemlan, E.~Nissimov and S.~Pacheva,
  ``$\Lambda$CDM as a Noether Symmetry in Cosmology,''
  Int.\ J.\ Mod.\ Phys.\ D {\bf 26}, 2050104 (2020), arXiv:2003.13146.
  \item   \textbf{D.~Benisty}, E.~I.~Guendelman, E.~Nissimov and S.~Pacheva,
  ``Quintessential Inflation with Dynamical Higgs  Generation as an Affine Gravity,'' 
  Symmetry {\bf 12}, 734 (2020), arXiv: 2003.04723.
  \item   \textbf{D.~Benisty}, E.~Guendelman, E.~Nissimov and S.~Pacheva,
   ``Dynamically Generated Inflationary $\Lambda$CDM,''
  Symmetry {\bf 12}, no. 3, 481 (2020), 
  arXiv: 2002.04110.
    \item  \textbf{D.~Benisty}, E.~I.~Guendelman and E.~N.~Saridakis,
  ``The Scale Factor Potential Approach to Inflation,''
  Eur.\ Phys.\ J.\ C {\bf 80}, no. 5, 480 (2020), arXiv: 1909.01982.
  \item  \textbf{D.~Benisty}, E.~I.~Guendelman, E.~Nissimov and S.~Pacheva,
  ``Dynamically generated inflationary two-field potential via non-Riemannian volume forms,''
  Nucl.\ Phys.\ B {\bf 951}, 114907 (2020), arXiv: 1907.07625
  \item   \textbf{D.~Benisty}, E.~Guendelman, E.~Nissimov and S.~Pacheva,
  ``Dynamically Generated Inflation from Non-Riemannian Volume Forms,''
  Eur.\ Phys.\ J.\ C {\bf 79}, no. 9, 806 (2019),
  arXiv: 1906.06691.
  \item   \textbf{D.~Benisty}, E.~I.~Guendelman, E.~N.~Saridakis, H.~Stoecker, J.~Struckmeier and D.~Vasak,
  ``Inflation from fermions with curvature-dependent mass,''
  Phys.\ Rev.\ D {\bf 100}, no. 4, 043523 (2019), arXiv: 1905.03731.
 \item   \textbf{D.~Benisty}, E.~I.~Guendelman,
  ``Cosmological Principle in Newtonian Dynamics,''
  Mod.\ Phys.\ Lett.\ A {\bf 35}, no. 16, 2050131 (2020),
  arXiv: 1902.06511.
\item \textbf{D.~Benisty}, E.~I.~Guendelman, D.~Vasak, J.~Struckmeier and H.~Stoecker,
  ``Quadratic curvature theories formulated as Covariant Canonical Gauge theories of Gravity,''
  Phys.\ Rev.\ D {\bf 98}, no. 10, 106021 (2018), arXiv: 1809.10447.
  \item  \textbf{D.~Benisty}, E.~I.~Guendelman,
  ``Two scalar fields inflation from scale-invariant gravity with modified measure,''
  Class.\ Quant.\ Grav.\  {\bf 36}, no. 9, 095001 (2019),
  arXiv: 1809.09866
  \item   \textbf{D.~Benisty}, D.~Vasak, E.~Guendelman and J.~Struckmeier,
  ``Energy transfer from spacetime into matter and a bouncing inflation from covariant canonical gauge theory of gravity,''
  Mod.\ Phys.\ Lett.\ A {\bf 34}, no. 21, 1950164 (2019),
  arXiv: 1807.03557
  \item   \textbf{D.~Benisty}, E.~I.~Guendelman,
  ``Correspondence between the first and second order formalism by a metricity constraint,''
  Phys.\ Rev.\ D {\bf 98}, no. 4, 044023 (2018), arXiv: 1805.09667
  \item  S.~Bahamonde, \textbf{D.~Benisty}, E.~I.~Guendelman,
  ``Linear potentials in galaxy halos by Asymmetric Wormholes,''
  Universe {\bf 4}, no. 11, 112 (2018), arXiv: 1801.08334
  \item    \textbf{D.~Benisty}, E.~I.~Guendelman,
  ``A transition between bouncing hyper-inflation to $\Lambda$CDM from diffusive scalar fields,''
  Int.\ J.\ Mod.\ Phys.\ A {\bf 33}, no. 20, 1850119 (2018), arXiv: 1710.10588.
  \item    \textbf{D.~Benisty}, E.~I.~Guendelman,
  ``Unified DE–DM with diffusive interactions scenario from scalar fields,''
  Int.\ J.\ Mod.\ Phys.\ D {\bf 26}, no. 12, 1743021 (2017).
\end{enumerate}
Conference proceedings: 
\begin{enumerate}
  \item  \textbf{D.~Benisty}, E.~I.~Guendelman and J.~Struckmeier,
``Gauge Theory of Gravity Based on the Correspondence Between the $1^{st}$ and the $2^{nd}$ Order Formalisms,''
Springer Proc. Math. Stat. \textbf{335} (2019), 309-316, arXiv: 1808.01978.
\item \textbf{D.~Benisty}, E.~Guendelman, A.~Kaganovich, E.~Nissimov and S.~Pacheva,
``Modified Gravity Theories Based on the Non-canonical Volume-Form Formalism,''
Springer Proc. Math. Stat. \textbf{335} (2019), 239-252, arXiv: 1905.09933.

\end{enumerate}
\mbox{} 
\newpage

\section{Acknowledgements}

First of all, I would like to thank my supervisor, Professor Eduardo Guendelman, for his great support throughout the research. His guidance and advice are beyond the standard and will never be forgotten. In addition, I would like to thank my parents and my family for their strong love, encouragement and support that enabled me to do my research. and friends for their encouragement. Finally, I would like to express my gratitude to Hanna Nahe, who gave me guidance and directions from her wisdom during my PhD studies.

I wish to thank for Professor Horst Stoecker, Professor Jurgen Struckmeier, Dr. David Vasak and his group, for the opportunities they opened for me during my studies. I thank to Mrs. Margarethe Puschmann and the Herbert Puschmann Stiftung for Frankfurt Institute For Advanced Studies in the Verein der Freunde and Foerderer der Goethe University. I thank to Miss Malka Elsje Shamir-de Leeuw and Dr. Rechali Cohen for editing. I thank for the antonymous referee for his/her comments and revisions. 

David Benisty

\newpage													
\mbox{} 
\newpage
\section{Introduction} 
The best explanation for the accelerated expansion of our  {Universe is the $\Lambda$CDM model, which addresses dark energy ($\Lambda$) and Cold Dark Matter (CDM) components} \cite{Perlmutter:1998np,Riess:1998cb,Angus:2018tko,Zhang:2018gbq,Lambda-CDM-1,Sandage:2006cv,Riess:2019cxk,RevModPhys.61.1}.   {Some models claim that the dark energy is quantum fluctuations of the vacuum, while dark matter is an additional substance that doesn't interact with light.} Yet, the true nature of these two phenomena is still a mystery  {raising} fundamental problems \cite{Verde:2019ivm}: The cosmological constant problem discusses large disagreement between the vacuum expectation value of the energy momentum tensor which comes from quantum field theory on one side, and the observed value of the dark energy density on the other side \cite{Weinberg:1988cp,Martin:2012bt,Padilla:2015aaa,Benisty:2019fzt,Lombriser:2019jia}. This problem is considered to be one from the biggest problems in modern physics. Another fundamental problem is the coincidence between the ratio of dark matter and dark energy in our  {Universe}. In order to solve this problem many approaches emerged \cite{Velten:2014nra}. Those models introduce some interaction between dark energy and dark matter \cite{Dam:2019prv,Kisslinger:2019ysx,Kamada:2019jch}. Some models claim that the vacuum  {energy} is running with the evolution of the  {Universe} and even may  {give} better data fitting then the usual $\Lambda$CDM model \cite{Sola:2016zeg,Sola:2016jky,Saridakis:2019qwt,DiValentino:2020srs,DiValentino:2020vvd,DiValentino:2020zio,DiValentino:2020vhf,Abadi:2020hbr}. 

 {Interaction between dark matter and dark energy was considered in many cases,} such as \cite{DiValentino:2019jae,Arevalo:2016epc,Anagnostopoulos:2017iao,Saridakis:2018unr}. Unification between dark energy and dark matter from an action principle were obtained from scalar fields \cite{Scherrer:2004au,Arbey:2006it} or by other models \cite{Chen:2008ft,Leon:2013qh,Leon:2012mt,Cai:2015emx,Kofinas:2014aka,Skugoreva:2014ena,Bahamonde:2015zma,Bahamonde:2015hza}. Others describe the dark matter as an effective scalar field \cite{Khalifeh:2019zfi,Solomon:2019qgf,Ganz:2018vzg,Gorji:2018okn,Arroja:2017msd}.

  {One has to take into consideration the  Neutron Stars (NS) merger measurement on 17 August 2017 the \cite{GBM:2017lvd}. The observation detects the Gravitational Waves from the NS merger and its associated electromagnetic counterparts. Combined analysis of the signals yields an equivalence between the speed of gravity and the speed of light. The equivalence invalidates many modifications to General theory Relativity \cite{Lombriser:2016yzn,Creminelli:2017sry,Ezquiaga:2017ekz}. }

One elementary way to parametrize dark energy is by a scalar field $\phi$ - i.e.  {quintessence models} \cite{Ratra:1987rm,Caldwell:1997ii,Zlatev:1998tr,Chiba:1999ka,Barreiro:1999zs,Bento:2002ps,dePutter:2007ny,Tsujikawa:2013fta,Babichev:2018twg,Kehayias:2019gir,Oikonomou:2019muq,Chakraborty:2019swx,Chervon:2019sey}.   {A flat quintessential potential does not give a dark matter component.} Therefore \cite{Gao:2009me} uses a scalar field model that unifies dark energy and dark matter from one scalar field. However, the model lacks of an action principle. This thesis introduces the action principle based on the model \cite{Gao:2009me} with  {DST} \cite{Guendelman:2009ck}. This generalization reduces to regular dark energy and dark matter components in particular cases, and addresses some extensions. These extensions fit with the late time accelerated expansion of the  {Universe}. Before we present the DST, we will summarize the foundations of the standard model of cosmology.

\subsection{FRW cosmology}
\label{FRW}
The dynamics of the  {Universe} is described by the Einstein equations which are the variation of the Einstein-Hilbert action. A simple way to solve the Einstein equations uses generic symmetries. The Friedman-Lemaitre-Robertson-Walker (FLRW) metric \cite{Friedman:1922kd} is the standard ansatz in cosmology, which is based on the assumption of a homogeneous and isotropic  {Universe} at any point ("The cosmological principle").  {The FLRW metric reads:}
\begin{equation}\label{frwmet}
d s^{2}=-dt^{2}+a^{2}(t)\left[\frac{d r^{2}}{1-Kr^{2}}+r^{2}\left(d\theta^{2}+\sin^{2}\theta d\phi^{2}\right)\right].
\end{equation}
In this problem the dynamics is associated with the scale factor - $a$. Einstein equations allow us to determine
the scale factor provided the matter content of the  {Universe} is specified. The 
constant $K$ in the metric (\ref{frwmet}) describes the geometry of the spatial section of spacetime, with closed, flat and open  {Universe}s corresponding to $K=+1, 0, -1$, respectively.

The differential equations for the scale factor and the matter density follow from Einstein's equations:
\begin{equation}
G^{\mu\nu} = R^{\mu\nu}-\frac{1}{2}
g^{\mu\nu}R=\frac{8 \pi G}{c^4} T^{\mu\nu}\,,
\label{Einsteineq}
\end{equation}
 {where $R^{\mu}_{\nu}$ is the Ricci tensor, $R$ is the Ricci scalar, $G^{\mu}_{\nu}$ is the Einstein tensor and $T^{\mu}_{\nu}$ is the energy momentum tensor.} In the FRW background (\ref{frwmet}) the curvature terms are given by:
\begin{eqnarray}
R_{0}^0 &=& \frac{3\ddot{a}}{a}\,, \\
R^{i}_{j}&=& \left(\frac{\ddot{a}}{a}+\frac{2\dot{a}^2}{a^2}+
\frac{2K}{a^2} \right) \delta^{i}_{j}\,, \\
R&=&6 \left(\frac{\ddot{a}}{a}+\frac{\dot{a}^2}{a^2}
+\frac{K}{a^2}\right)\,,
\end{eqnarray}
where the "dot" denotes a derivative with respect to the cosmic time and the indexes $i,j$ refers to $1,2,3$. To set up the source term of the energy momentum tensor, the  {Universe} is usually modeled as a perfect fluid. The appropriate energy-momentum tensor is then
\begin{equation}\label{eq:robwalk-T}
T^{\alpha}_{\xi}=\mathrm{diag}(\rho,-p,-p,-p),
\end{equation}
 {where we set $c = 1$}.   {Due to the symmetry properties, the density and the pressure are time dependent. The Friedmann equations read:} 
\begin{eqnarray}
 \label{HubbleeqI}
&& H^2 \equiv \left(\frac{\dot{a}}{a}\right)^2
=\frac {8\pi G \rho}{3}-\frac {K}{a^2}\,, \\
\label{dotHeq}
&&\dot{H}=-4\pi G(p+\rho)+\frac{K}{a^2}\,,
\end{eqnarray}
where $H$ is the Hubble parameter. The energy momentum tensor is conserved by virtue of the Bianchi identities, leading to the continuity equation
\begin{equation}
\dot{\rho}+3 H(\rho+p)=0\,.
\label{conteq}
\end{equation}
Equation (\ref{conteq}) can be derived from 
Eqs.~(\ref{HubbleeqI}) and (\ref{dotHeq}), 
 {which means that only two of} Eqs.~(\ref{HubbleeqI}), (\ref{dotHeq}) and
(\ref{conteq}) are independent. Eliminating the $K/a^2$ term 
from Eqs.~(\ref{HubbleeqI}) and (\ref{dotHeq}), we obtain
\begin{eqnarray}
\label{acceleq}
\frac{\ddot{a}}{a}=
-\frac {4 \pi G}{3} \left(\rho+3p\right)\,.
\end{eqnarray}
 {Hence accelerated expansion occurs for $\rho+3p<0$.} One can rewrite Eq.~(\ref{HubbleeqI}) in the form of $\Omega-1= K/(aH)^2$, where $\Omega$ is the fraction between the density $\rho(t)$ and the critical density $\rho_c(t)=3H^2(t)/8 \pi G$.  {Observations show that the current  {Universe} has a spatially flat geometry with $K = 0$ \cite{Aghanim:2018eyx}. }
\subsection{Constant Equation of State}
The density $\rho(t)$ and the pressure $p(t)$ are related via an \emph{equation of state}, which, 
for a perfect fluid, is characterized by a constant parameter $\omega$:
\begin{equation}\label{eq:eos}
\omega = p/\rho,
\end{equation}
where $w$ in our simple model is assumed to be a constant. 
Then by solving the Einstein equations given in Eqs.~(\ref{HubbleeqI})
and (\ref{dotHeq}) with $K=0$, the Friedmann equations yield:
\begin{subequations}
\begin{equation}
a \propto (t-t_{0})^{\frac{2}{3(1+w)}},    
\end{equation}
\begin{equation}
\rho \propto a^{-3(1+w)}, 
\end{equation}
\end{subequations}
 {where $t_{0}$ is constant. Notice that the above solution is valid for $w \neq -1$. The case of $w = -1$ gives the \emph{de-Sitter Space} solution:}
\begin{subequations}
\begin{equation}
a \propto e^{H_0 t}  , 
\end{equation}
\begin{equation}
\rho = Const,
\end{equation}
\end{subequations}
with a constant Hubble parameter.   {This solution is apparently the solution for the early and late stages of our  {Universe}. The inflationary paradigm is considered to be a necessary part of the  {Universe}, since it provides a solution to the the horizon, the flatness, and the monopole problems
\cite{Guth:1980zm,Guth:1982ec,Starobinsky:1979ty,Kazanas:1980tx,Starobinsky:1980te,Linde:1981mu,Albrecht:1982wi,Barrow:1983rx,Blau:1986cw}. Inflation predicts a short exponential expansion before the reheating the  {Universe}, whereas the late dark energy domination epoch explains the accelerated expansion of the current period of our  {Universe} \cite{Perlmutter:1998np,Weinberg:1988cp,Lombriser:2019jia,Copeland:2006wr,Frieman:2008sn,Riess:2019cxk}. Notably, the early inflationary energy density must be much larger than the late time energy density.  }

Beside this particular solution, there are two important solutions, that during the evolution of our  {Universe} appear to be a good approximation. The radiation dominated  {Universe} corresponds to $w=1/3$, whereas
the dust dominated  {Universe} to $w=0$. 
In those cases, the density read:
\begin{eqnarray}
{\rm Radiation}:~~a \propto (t-t_{0})^{1/2}\,,~~~
\rho \propto a^{-4}\,, \\
{\rm Dust}:~~a \propto (t-t_{0})^{2/3}\,,
~~~\rho \propto a^{-3}\,.
\end{eqnarray}
Both cases correspond to a decelerated expansion of the  {Universe}. In order to explain the accelerated expansion of our  {Universe}, we require an exotic vacuum energy, called ``dark energy'' with equation of state of $w = -1$. This exponential expansion arises by including only the cosmological constant $\Lambda$. The observations which constrain the value of the equation of state today to be close to $-1$  {\cite{Abbott:2017wau,Aghanim:2018eyx,Abbott:2018xao,Huang:2020tpm}.}, say relatively little about the time evolution of $w$, and so we can broaden our horizons and consider a situation in which the equation of state of dark energy changes with time. Scalar fields naturally arise in particle physics including string theory and these can act as candidates for dark energy. So far a wide variety of scalar-field dark energy models have been proposed.  {The simplest one is the quintessence model \cite{Peebles:1987ek,Ratra:1987rm,Wetterich:1994bg,Frieman:1995pm,Ferreira:1997au,Viana:1997mt,Copeland:1997et,Caldwell:1997ii,Zlatev:1998tr,Martin:2008qp,Tsujikawa:2013fta}.}

\newpage 
\subsection{Cosmology with matter Scalar fields}
\label{scalarmodel}
Quintessence is described by an ordinary scalar field $\phi$ minimally coupled to  {gravity}:
\begin{eqnarray}
S=\int {\rm d}^4x \sqrt{-g} \left[\frac{1}{16 \pi G} \mathcal{R} -\frac12 
g^{\mu\nu}\partial_\mu \phi \partial_\nu \phi-V(\phi) \right]\,, 
\label{NG}
\end{eqnarray}
where $V(\phi)$ is the potential of the field. From the variation with respect to the scalar field, we obtain the Klein Gordon equation:
\begin{equation}
 \Box \phi - V'(\phi) =  \frac{1}{\sqrt{-g}}\partial_\mu \left(g^{\mu\nu}\sqrt{-g}\partial_\nu \phi\right) - V'(\phi) = 0
\label{KG}
\end{equation}
  {Notice that the d'Alembertian here $\Box\phi$ is with respect to a particular metric and not for a flat spacetime. In a flat FLRW metric the variation (\ref{KG}) gives: }
\begin{eqnarray}
\ddot{\phi}+3H \dot{\phi}+
\frac{d V}{d \phi}=0\,.
\label{phieqF}
\end{eqnarray}
The energy momentum tensor of the field is derived by varying the action (\ref{NG})  {with respect to} $g^{\mu \nu}$:
\begin{equation}
T_{\mu\nu}=-\frac{2}{\sqrt{-g}}\frac{\delta S}{\delta
g^{\mu\nu}}\,
=\, \partial_{\mu} \phi \partial_{\nu} \phi-g_{\mu \nu} 
\left[{1 \over 2}g^{\alpha \beta} \partial_{\alpha} 
\phi \partial_{\beta} \phi+V(\phi)\right]\,.
\end{equation}
In the flat Friedmann background we obtain the energy density and pressure density of the scalar field:
\begin{equation}
\rho=T_0^0=\frac12 \dot{\phi}^2+V(\phi)\,,~~~
p=-T_i^i=\frac12 \dot{\phi}^2-V(\phi)\,.
\end{equation}
The Einstein equations yield:
\begin{eqnarray}
\label{H2}
& &H^2=\frac{8\pi G}{3} \left[\frac12 \dot{\phi}^2
+V(\phi) \right]\,, \\
\label{ddota}
& &\frac{\ddot{a}}{a}=-\frac{8\pi G}{3} 
\left[\dot{\phi}^2-V(\phi) \right]\,.
\end{eqnarray}
Notice that the continuity equation (\ref{conteq}) is derived by combining these equations. The equation of state for the field $\phi$ ranges in 
the region $-1 \le w_{\phi} \le 1$.
The slow-roll limit, $\dot{\phi}^2 \ll V(\phi)$, corresponds to $w_{\phi}=-1$,  {thus asymptomatically one has dark energy/ Cosmological Constant behavior. }   {In the case of a constant potential, stiff matter is obtained and the energy density evolves as $\rho \propto a^{-6}$.} In order to obtain dark matter behavior from the same scalar field,  {reference} \cite{Gao:2009me} extended the idea of the quintessence model.

\subsection{Unification of dark matter and dark energy}

 {Reference \cite{Gao:2009me} studies the consequences of modeling dark energy using a scalar field that is of non-Lagrangian type in order to address the coincidence problem. The model offers a unified description of dark energy and dark matter from a single scalar field.}

The model explores unification of dark matter and dark energy by direct insertion of a kinetic term into the energy momentum tensor.  This scalar is different from quintessence, having an equation of state between $-1$ and $0$. To remedy this it is necessary to incorporate a dynamical term, depending on $\nabla^{\mu}\phi$, into the equations. For quintessence this is done by including a canonical kinetic term in the Lagrangian; $\Lambda(\phi)$ then becomes the scalar field potential and the total dark energy density includes both potential and kinetic terms. Here we propose the simplest possible alternative, which is the direct insertion of a kinetic term into the energy momentum tensor:
\begin{eqnarray}
\frac{1}{\kappa^2}G_{\mu\nu}=
\Lambda\left(\phi\right)g_{\mu\nu}-
\frac{1}{2}\nabla_{\mu}\phi\nabla_{\nu}\phi
\;.
\end{eqnarray}
Now $\Lambda(\phi)$ is not necessarily a constant. From the Einstein equations above we obtain the density and pressure
of our scalar
\begin{eqnarray}
\label{eq:dp}
 &&\rho_{\textrm{sca}}=\frac{1}{2}\dot{\phi}^2+\Lambda\left(\phi\right),
\nonumber\\&&
p_{\textrm{sca}}=-\Lambda\left(\phi\right).
\end{eqnarray}
From the expressions of density and pressure, we know quintessence has the  {equation of state} $-1\leq w_{\textrm{qui}}\leq 1$ for $\Lambda\geq0$, while the  {scalar field} has $-1\leq w_{\textrm{sca}}\leq 0$. From the conservation equation we then know that the density of quintessence scales in the range $a^{-6}$ to $a^{0}$, while for the  {scalar field} the range is restricted to $a^{-3}$ and $a^{0}$. This property suggests that the  {scalar field} may play the role of both dark matter (scaling approximately as $a^{-3}$) and dark energy (scaling approximately as $a^{0}$). From the conservation of the energy momentum tensor the model yields:
\begin{eqnarray}
\label{eq:eoms}
\ddot{\phi}+\frac{3}{2}H\dot{\phi}+\Lambda'(\phi)=0\;.
\end{eqnarray}
  {For a constant potential $\Lambda(\phi) = \Lambda$, the scalar field solution is $\dot{\phi} \sim a^{-3/2}$. Therefore, Eq. (\ref{eq:dp}) gives:}
\begin{eqnarray}
\label{eq:dp2}
 &&\rho_{\textrm{sca}}=\frac{C}{a^3}+\Lambda,
\nonumber\\&&
p_{\textrm{sca}}=-\Lambda.
\end{eqnarray}
  {where $C$ is an integration constant. The kinetic term gives the dark matter component and the potential gives the dark energy component. This is unlike the standard quintessence model, where the kinetic term gives a stiff equation of state.}

  {This model is not formulated from an action principle. In order to solve this problem we added the DST to the scalar field action. Hence, this theory introduces a vector field $\chi_\mu$ that changes the scalar field dynamics.}
											
\mbox{} 
\newpage
\section{The basis of DST}
\subsection{A toy model}
In order to understand the formulation of  {DST}, we start with the simplest possible example: The dynamics of a  {classical} particle in a potential $V(x)$ where the energy conservation is assumed from the beginning. The energy reads:
\be
 \frac{1}{2}m v^2 + V(x) = E = \text{Const},
\label{energyFunctional}
\ee
with the  {speed} $v=dx/dt$. The action we assume for this dynamics is:
\be
\mathcal{S} = \int dt\, \mathcal{L} = \int dt\, \frac{dB}{dt} \left(\frac{1}{2}m v^2 + V(x)\right) ,
\lab{action}
\ee
where the dynamical variable $B$ gives the conservation of energy from  {its} variation (\ref{energyFunctional}). One can differentiate (\ref{energyFunctional}) with respect to time:
\be
m \dot{v} = - V'(x),
\label{eqforv}
\ee
which is the Newton's Second Law. 

One can check complete consistency with the other variations. The variation with respect to $x$ gives:
\be
m\frac{d}{dt}\left( \frac{dB}{dt}\frac{dx}{dt}\right) = \frac{dV(x)}{dx} \frac{dB}{dt}.
\label{yetanothereq}
\ee
Since the action \rf{action} is time independent, there is a conservation property:
\be
\mathcal{H} = \frac{dB}{dt} p_a + \frac{dx}{dt}p_x - \mathcal{L}  = \frac{dx}{dt}p_x = m v^2 \frac{dB}{dt} = \text{Const},
\label{Hamiltonian}
\ee
so that 
\begin{equation}
\frac{dB}{dt} = C/v^2.
\label{eq:con}
\end{equation}
Eq. (\ref{eq:con}) with Eq. (\ref{yetanothereq}) yield:
\be
m\frac{d}{dt}\left( \frac{C}{v}\right) =V'(x) \frac{C}{v^2},
\lab{final eq.}
\ee
which gives (\ref{eqforv}) exactly.   {The conservation of energy functional is implemented by $B$ variation and the additional conservation law determines $B$. This is a classical non-relativistic action. Let's see how to generalize this idea into a relativistic and covariant theories.}

\subsection{The DST action}
  {One of the basic features in the standard approach to theories of gravity is the local conservation of an energy momentum tensor. For example, the conservation of energy is derived from the time translation invariance principle. The local conservation of energy momentum tensor can emerge from a variation of a certain vector field.}  {Reference} \cite{Guendelman:2009ck} considers a four dimensional case where conservation of a symmetric energy momentum tensor $T^{\mu\nu}_{(\chi)}$ is imposed by introducing a term in the action:
\begin{equation} \label{action}
	\mathcal{S}_{(\chi)}=\int d^{4}x\sqrt{-g} \, \chi_{\mu;\nu}T_{\left(\chi\right)}^{\mu\nu}
\end{equation}
  {where $ \chi_{\mu;\nu}=\partial_{\nu}\chi_{\mu}-\Gamma_{\mu\nu}^{\lambda}\chi_{\lambda}$ and $\Gamma_{\mu\nu}$ is the Affine-Connection. We use the metric formalism (or second order formalism) where the connection is assumed to be the Levi-Civita symbol}:
\begin{equation}
\Gamma{^{\rho}_{\mu\nu}} = \left\{ \genfrac{}{}{0pt}{}{\rho}{\mu \nu} \right\} = \frac{1}{2} g^{\rho\lambda} (g_{\lambda\mu,\nu}+g_{\lambda\nu,\mu}-g_{\mu\nu,\lambda}).
\end{equation}
The vector field $\chi_\mu$ is called a  {"Dynamical Spacetime vector"}, because the energy
density of $T^{\mu\nu}_{(\chi)}$ is a canonically conjugated variable to $\chi_0$, which is what we  {expect} from a  {Dynamical Time}:
\begin{equation} 
	\pi_{\chi_{0}} = \frac{\partial \mathcal{L}}{\partial \dot{\chi}^0} = T^{0}_{0} (\chi) 
\end{equation}
 {The variation with respect to $ \chi_{\mu} $ gives a covariant conservation law:}
\begin{equation}
\nabla_{\mu}T_{\left(\chi\right)}^{\mu\nu}=0
\end{equation}
From the variation of the action with respect to the metric, we get a conserved stress energy tensor $G^{\mu\nu}$ (in appropriate units), which is well known from Einstein's equation:
\begin{equation}
	G^{\mu\nu}=\frac{2}{\sqrt{-g}}\frac{\delta\sqrt{-g}}{\delta g^{\mu\nu}}[\mathcal{L}_{\chi}+\mathcal{L}_{m}]\,,\quad \nabla_{\mu} G^{\mu\nu}=0\,.
\end{equation}
where $G^{\mu\nu}$ is Einstein tensor, $\mathcal{L}_{\chi}$ is the Lagrangian in (\ref{action}) and $\mathcal{L}_{m}$ is an optional action that involve other contributions. Cosmological solutions with a scalar field behaving as radiation, in the context of gravitational theory with  {Dynamical Time} are discussed in \cite{Benisty:2016ybt}.

\subsection{Diffusive Action}
 {The} diffusion equation can be generalized into a  {curved spacetime} by defining a non-conserved stress energy tensor $T^{\mu\nu}$ with a current source $j^\mu$ \cite{Calogero:2011re,Haba:2009by}:
\begin{equation} \label{chitensor}
\nabla_\mu T^{\mu\nu}=3\sigma j^\nu
\end{equation}
where $\sigma$ is the diffusion coefficient of the fluid.
The current $j^\mu$ is a time-like covariant conserved vector field $j^{\mu}_{;\mu}=0$ which describe the conservation of the number of particles in the system. In order to break the conservation of $T^{\mu\nu}_{(\chi)}$,  {we use the action:}
\begin{equation} \label{nhd1}
S_{(\chi,A)}=\int d^{4}x\sqrt{-g}\chi_{\mu;\nu}T_{\left(\chi\right)}^{\mu\nu} +\frac{\sigma}{2}\int d^4x \sqrt{-g}(\chi_{\mu}+\partial_{\mu}A)^2 
\end{equation} 
 {where $A$ is a different scalar field from $\phi$.} From a variation with respect to the Dynamical Spacetime vector field $\chi_{\mu}$ we obtain:
\begin{equation} \label{nhd2}
\nabla_{\nu}T_{\left(\chi\right)}^{\mu\nu}=\sigma(\chi^{\mu}+\partial^{\mu}A)= f^\mu,
\end{equation} 
 {where $f^\mu=\sigma (\chi^{\mu}+\partial^{\mu}A)$ is a current source for the stress energy momentum tensor} $T_{\left(\chi\right)}^{\mu\nu}$.
From the variation with respect to the new scalar $A$, a covariant conservation of the  {current indeed emerges}: 
\begin{equation}\label{nhd3}
\nabla_{\mu}f^\mu=\nabla_{\mu}(\chi^{\mu}+\partial^{\mu}A)=0
\end{equation} 
Using different expressions for $T^{\mu\nu}_{(\chi)}$ which depends on different variables, will give the conditions between the Dynamical Spacetime vector field $\chi_\mu$ and the other variables.

A particular case of diffusive energy theories is obtained when $\sigma \to \infty$. In this case, the contribution of the current $f_\mu$ in the equations of motion goes to zero, and from this constraint the vector field becomes to a gradient of the scalar:
\begin{equation}
f_\mu=\sigma(\chi_{\mu}+\partial_{\mu}A) =0 \quad \Rightarrow \quad \chi_{\mu}=-\partial_{\mu}A
\end{equation}
The theory (\ref{nhd1}) changes to a theory with higher derivatives:
\begin{equation}\label{action2}
\mathcal{S}= - \int d^{4}x\sqrt{-g} \, T_{\left(\chi\right)}^{\mu\nu} \, \nabla_\mu \nabla_\nu A \,  \end{equation}
The variation with respect to the scalar $A$ gives $\nabla_\mu \nabla_\nu T^{\mu\nu}_{(\chi)}=0$  {which corresponds to} the variations (\ref{nhd2}-\ref{nhd3}). In the DST we obtain  {four} equations of motion from the variation of $\chi_\mu$, which corresponds to a covariant conservation of energy momentum tensor $\nabla_\mu T^{\mu\nu}_{(\chi)}=0$. By changing the generic  {four} vector to a gradient of a scalar $\partial_{\mu}\chi$, the number of conditions reduces from  {four} to  {one} and instead of the conservation of energy momentum tensor,  {we are left} with a covariant conservation of the current $f^{\nu}=\nabla_\mu T^{\mu\nu}_{(\chi)}$.

\subsection{Overview of the Thesis}
In this thesis we formulate the DST in the  {framework of cosmological solutions}, which produces  {a natural} unification of dark energy and dark matter. In the first paper "Unified dark energy and dark matter from dynamical spacetime" we  {formulate} the basics for the complete theory and demonstrate simple analytic solutions.   {The DST introduces a Lagrange
multiplier in addition to the quintessential scalar field, that forces the kinetic term to act as dark matter.} In the second paper - "Dark energy and dark matter unification from dynamical spacetime: observational  {constrain} and cosmological implications" we extend the solution for the dynamics between dark energy and dark matter, using different potentials and constraint additional parameters with data of the late time accelerated expansion. Including higher dimensions with different  {scale factors} for the same family of theories produces an inflationary scenario, as  {described} in the third paper - "Inflation compactification from dynamical spacetime", which implies a new suggestion for  {an inflationary} solution that wasn't considered yet. We extended the DST theory to a "diffusive theories" which  {allows} to the dark energy and dark matter components to exchange energy. In the last section we discuss the results. In the fourth paper - "Interacting Diffusive Unified Dark Energy and Dark Matter from Scalar Fields", we solve the complete analytic solution for the simplest version of the theory. In the fifth paper - "Unification of dark energy and dark matter from diffusive cosmology", we analyze the complete combination of the stress energy momentum tensor with regards to diffusive interactions and we show analytically and numerically that all of the solutions  {yield} to the stable $\Lambda$CDM model with new families of solutions. In the last section we discuss the results.

\newpage													
\mbox{} 
\newpage
\section{The foundations of the model}
\label{sec:basis}
  \textbf{D.~Benisty}  and E.~I.~Guendelman,
  
``Unified dark energy and dark matter from dynamical spacetime",

Phys.\ Rev.\ D {\bf 98}, no. 2, 023506 (2018)

  {This paper unifies dark energy and dark matter from one scalar field through an action principle. Introducing the coupling of a Dynamical Spacetime vector field to the energy momentum tensor from \cite{Gao:2009me} gives the $\Lambda$CDM model with possible bouncing equation of state.} For the $\Lambda$CDM solution without the bouncing solution, the Dynamical Spacetime vector is equivalent to the cosmic time.
										
\includepdf[pages=-]{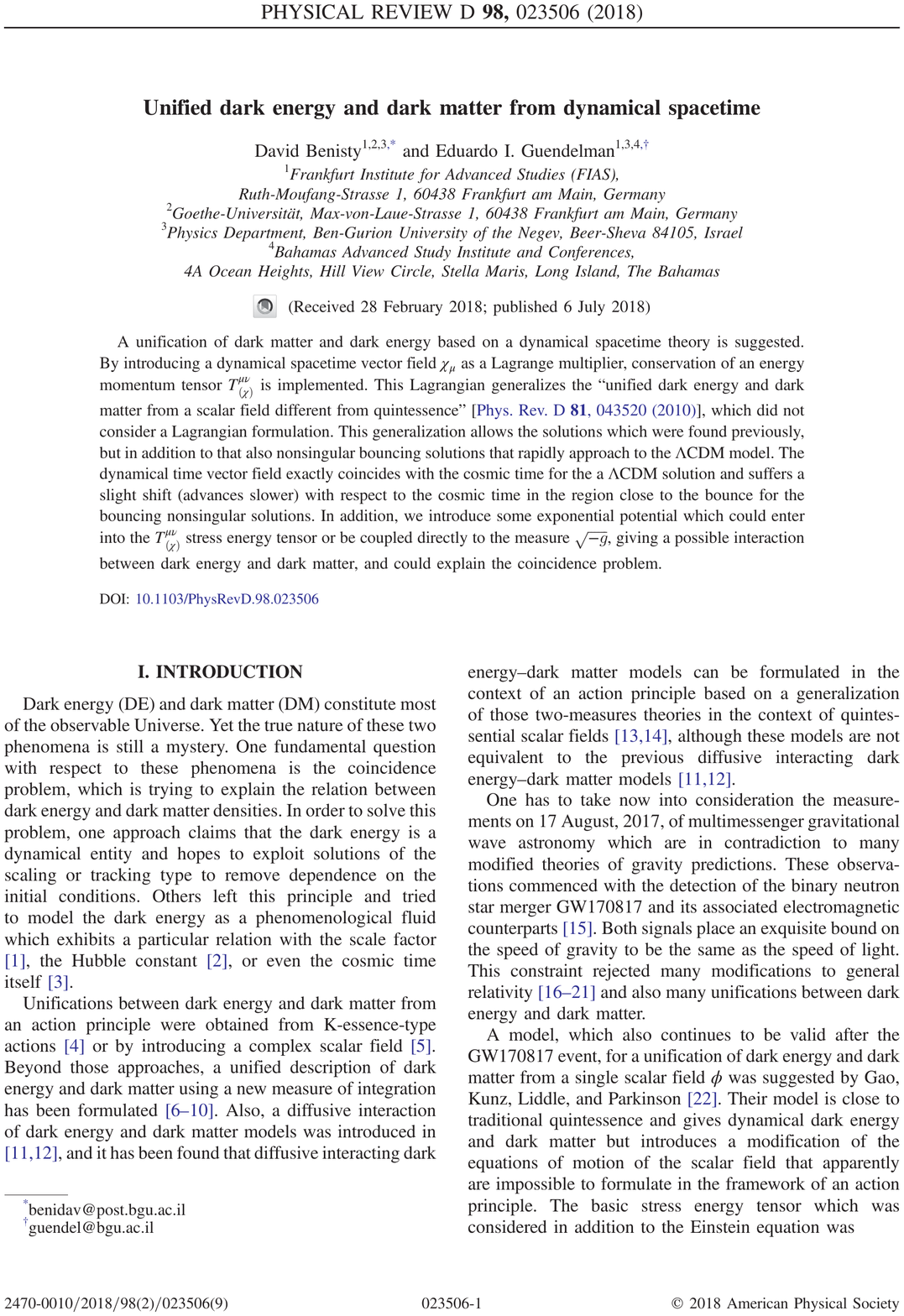}

\newpage													
\mbox{} 
\newpage 

\section{Observational constraints}
  F.~K.~Anagnostopoulos, \textbf{D.~Benisty} , S.~Basilakos and E.~I.~Guendelman,

  ``Dark energy and dark matter unification from dynamical spacetime: observational constraints and cosmological implications,''

JCAP {\bf 1906}, no. 06, 003 (2019)

  {This paper explores few potentials with theoretical and observational constraints. The Autonomous System Method revels stable late-time attractors. For a flat potential, the Hubble constant is in agreement with Planck 2018 results (within $\sim 1 \sigma$).}
										
\includepdf[pages=-]{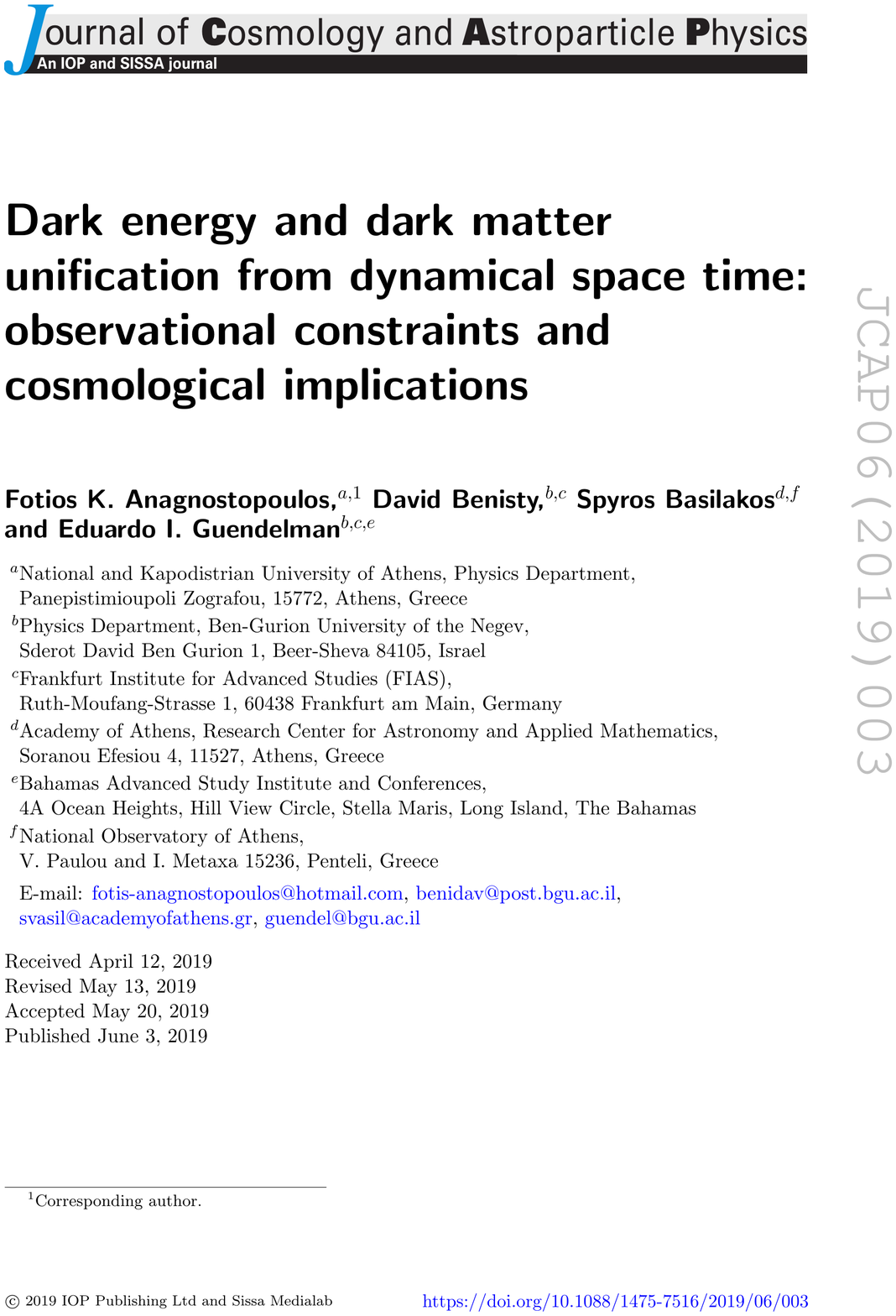}
\newpage													
\mbox{} 
\newpage

\section{Inflation from higher dimensions}
\textbf{D.~Benisty} and E.~I.~Guendelman,
  
``Inflation compactification from dynamical spacetime",   
  
Phys.\ Rev.\ D {\bf 98}, no. 4, 043522 (2018)

  {This article studies the basics of inflation compactification mechanism with the DST. The unification of dark energy, dark matter with the additional bounce of the volume naturally prevents the collapse of the  {Universe and gives a lower bound for the volume}. Consequently, for some values of the anisotropy parameter $E$, the total volume oscillates, the ordinary dimensions increase exponentially with an oscillatory modulation and the extra dimensions decrease correspondingly.} 
\includepdf[pages=-]{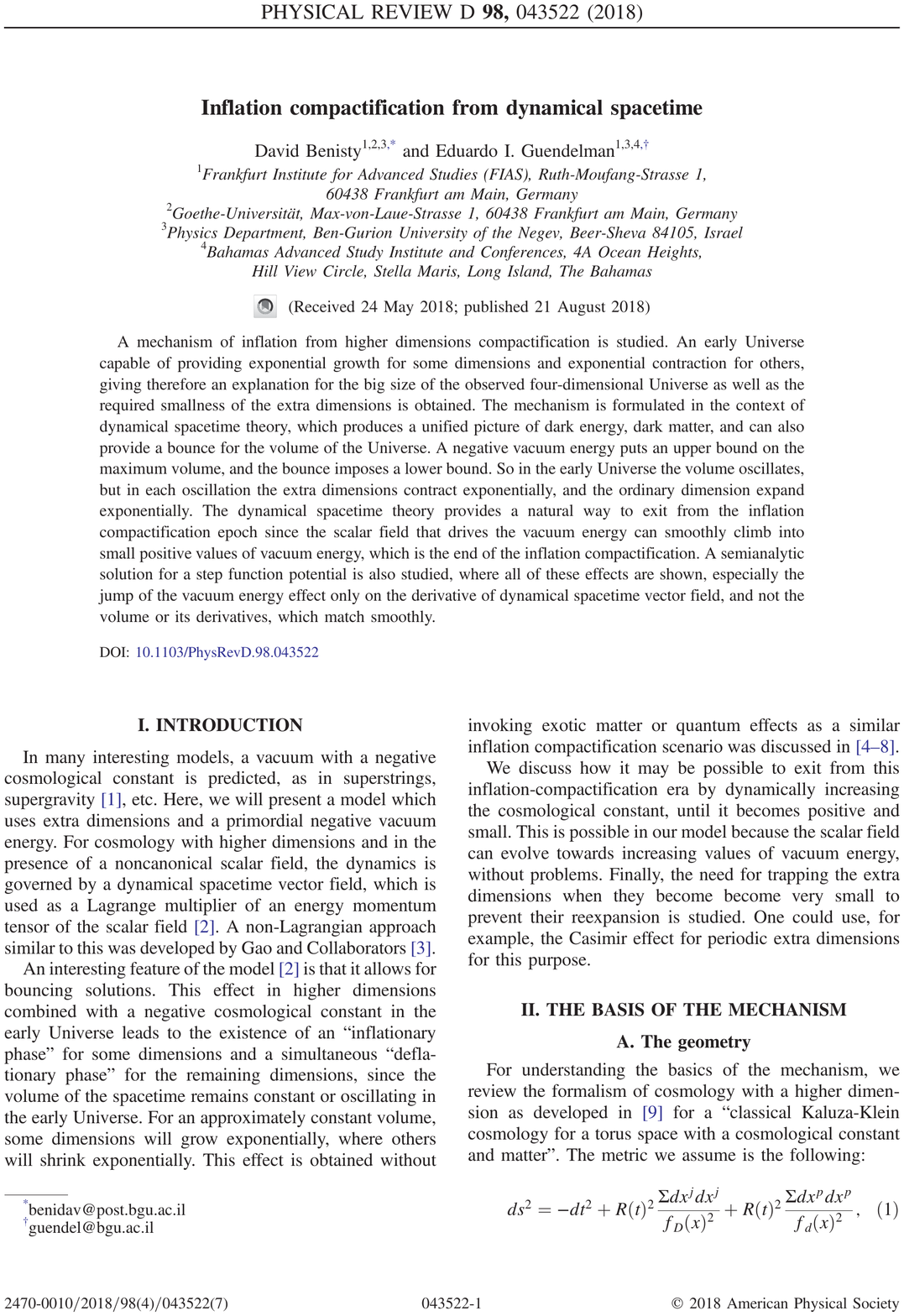}
\newpage

\mbox{} 
\newpage
\section{Diffusion effects from extended DST}
  \textbf{D.~Benisty} and E.~I.~Guendelman,

``Interacting Diffusive Unified Dark Energy and Dark Matter from Scalar Fields", 

Eur.\ Phys.\ J.\ C {\bf 77}, no. 6, 396 (2017)

  {This paper generalizes the DST by replacing the Dynamical Spacetime vector $\chi_\mu$ into a gradient $\partial_{\mu}\chi$. The replacement introduces a  current source for the energy momentum tensor $T^{\mu\nu}_{(\chi)}$. The current dissipates in an expanding  {Universe} and asymptotically the energy momentum tensor is conserved. The new mechanism predicts an energy transfer between the dark energy and dark matter components.}
\includepdf[pages=-]{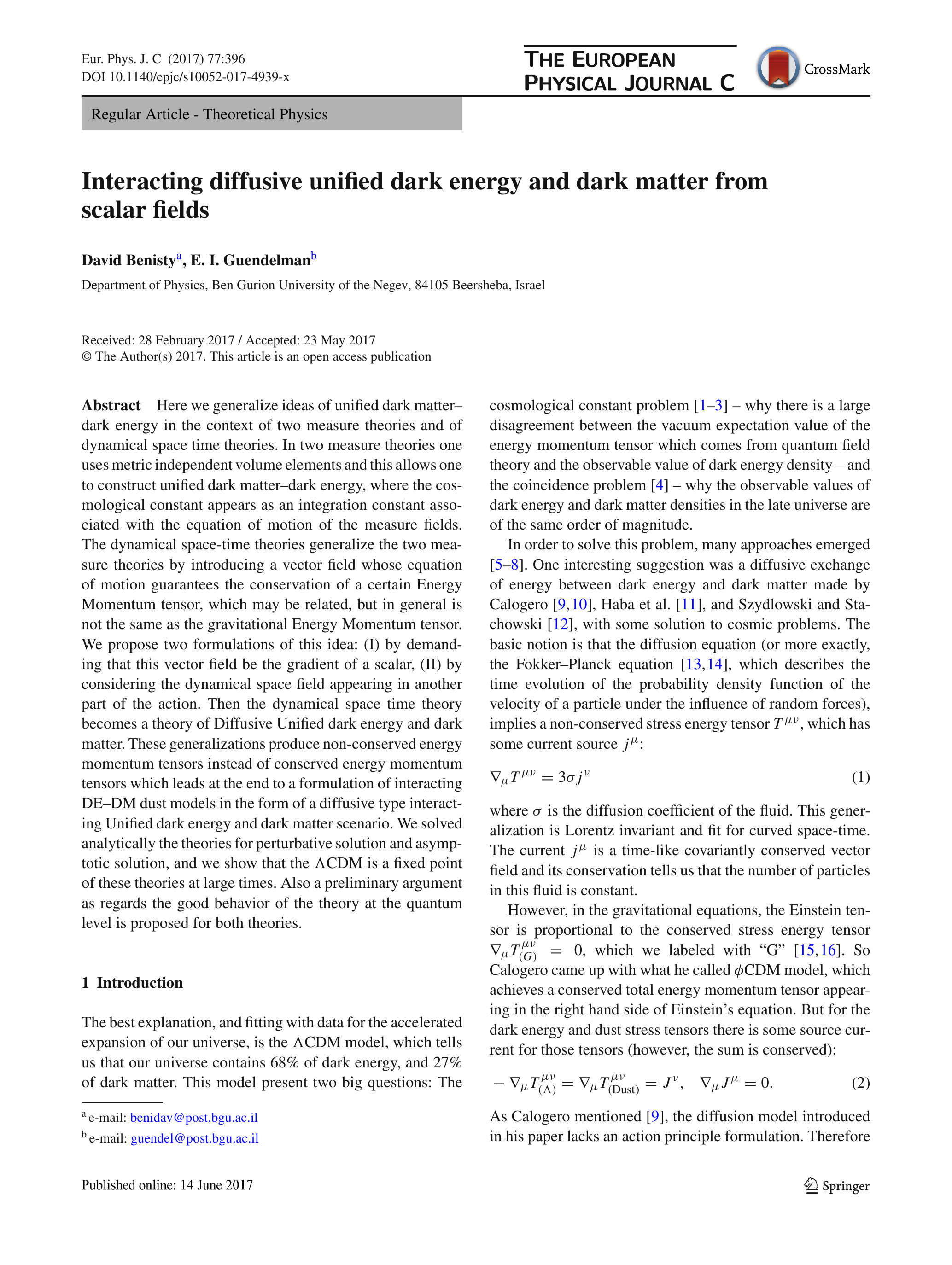}
\newpage													
\mbox{} 
\newpage
\section{General consideration with diffusion effects}
\label{sec:DE}
 D.~Benisty, E.~Guendelman and Z.~Haba,

``Unification of dark energy and dark matter from diffusive cosmology"

Phys.\ Rev.\ D {\bf 99}, no. 12, 123521 (2019)

  {This paper tests the general combination of $T^{\mu\nu}_{(\chi)}$ using the $\lambda_1$ and $\lambda_2$ parameters. This formulation of dark energy and dark matter  {has a direct} correspondence with the behavior  {of the non-Lagrangian} formulations of the dark energy and dark matter interactions only in the case $\lambda_2 = 0 $. In the other cases, the asymptotic behavior is different and the Dynamical Time approaches $1/3H_0$ asymptotically. All of the solutions have an asymptotically stable $\Lambda$CDM behavior.}
\includepdf[pages=-]{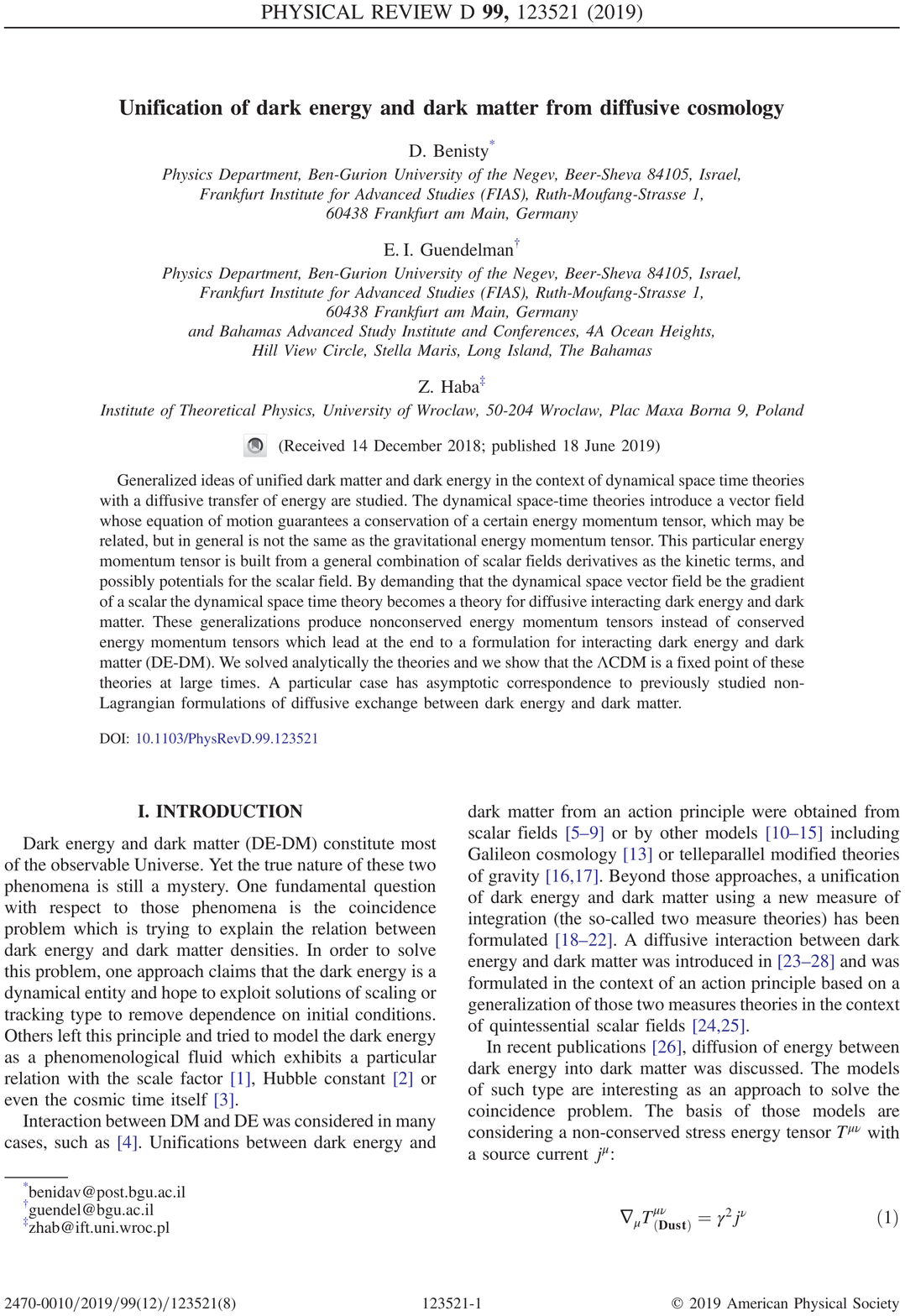}
\newpage
\mbox{} 
\newpage

\section{Discussion}
This thesis is based on the papers: \cite{Benisty:2018qed,Anagnostopoulos:2019myt,Benisty:2018gzx,Benisty:2017eqh,Benisty:2018oyy}.   {We formulate the DST and solve it for cosmological solutions, yielding a unification of dark energy and dark matter. The DST introduces a Lagrange multiplier that forces the kinetic part to mimic a dark matter component.}

The scalar field model of dark energy uses the slow roll assumption  {in order to get} effective dark energy behavior. The potential is more dominant  {than} the kinetic part $\dot{\phi}^2 \ll V(\phi)$. For the case of a constant potential the kinetic term gives a  {stiff} equation of state ($\rho = p$) to the matter fields, with a contribution of $\rho_\phi \sim 1/a^6$ to the Friedmann equation. From the measurements of the late time expansion, we don't see that component. Therefore, \cite{Gao:2009me} formulates a modified energy momentum tensor. The modified energy momentum tensor yields a dark matter contribution ($p = 0$) for a constant potential. This elegant model produces an interaction between dark energy and dark matter components with a corresponding potential. However, this model lacks an action principle.

In \cite{Benisty:2018qed} we  {used the DST} to formulate the unification of dark energy and the dark matter components. The energy momentum tensor that  {introduced in the DST is the same one that was introduced} in \cite{Gao:2009me}. From the variation with respect to the Dynamical Time Vector, the conservation of the energy momentum tensor $T^{\mu\nu}_{(\chi)}$ forces the kinetic term of the scalar field to  {behaves} as a dark matter component $\rho_\phi \sim 1/a^3$. In many situations the solutions studied in \cite{Gao:2009me} can be also obtained here. However, there are other solutions  {such as non-singular bounce solutions} which are not present in \cite{Gao:2009me}.

For the constant potentials the  {Dynamical Time} corresponds to the cosmic time. In particular, for exact $\Lambda$CDM solution the cosmic time and the  {Dynamical Time} coincide. The solution yields the $\Lambda$CDM background with a bouncing solution that solves the singularity problem. For  {an exponential} potential, the solution behaves differently for the early  {Universe}, but approaches $\Lambda$CDM for the late  {Universe}. Moreover,  \cite{Benisty:2018qed} proves the stability of the constant and the exponential potential.

 {Reference} \cite{Anagnostopoulos:2019myt} investigates many potentials with constraints from the late  {Universe} data. In this work we  {performed} the  {autonomous system method and showed} that under certain circumstances the DST includes stable late-time attractors. The asymptotic solution approaches $\Lambda$CDM model for those potentials. The observational constraints regarding the Hubble constant are in agreement (within $\sim 1 \sigma$) with those of Planck \cite{Aghanim:2019ame}. In addition, the results are compatible at $\sim 2\sigma$ level with the $H_0$ measurement obtained from Cepheids  {\cite{Freedman:2000cf,Riess:1998cb,Perlmutter:1998np,Riess:2016jrr,Riess:2020fzl}}. On top of that \cite{Anagnostopoulos:2019myt} finds that one of the  {models} with constant potentials have the smallest deviation from $\Lambda$CDM, where the confidence level is close to $\sim 1.5\sigma$. In addition, \cite{Anagnostopoulos:2019myt} explicitly  {checked} the compatibility of DST within the standard BBN using the average bound on the possible variation of the BBN speed-up factor.  {Reference} \cite{Anagnostopoulos:2019myt} shows that the deviation from the Hubble rate of $\Lambda$CDM for the radiation dominant era is not larger then $10 \% $. Therefore, the BBN production is still applicable with those potentials.

 {Reference} \cite{Benisty:2018gzx} studied the solution of DST with higher dimensions.  {Reference} \cite{Benisty:2018gzx} studied inflation solutions from the interplay of ordinary and higher dimensions. In the case of isotropic pressure, the solution is obtained from the total volume without dependence on an individual scale factor. The evolution for the total volume is calculated directly for the anisotropy constant $E$. The DST naturally prevents the collapse of the  {Universe} from the additional bounce and obtains a lower bound for the volume of the  {Universe}. Likewise the presence of a negative cosmological constant prevents the volume from become very large in the early  {Universe}.

  {There is an effective potential that governs the evolution of the total volume. In the case the effective potential includes a minimum, the total volume oscillates. Consequently, the ordinary dimensions increase and the extra dimensions decrease. The DST provides a natural way to exit from the inflation compactification epoch by one potential, which drives directly the evolution of the scalar field and a different potential which determines the value of the dark energy. As a consequence, the scalar field may drive the vacuum energy smoothly to rise into a small positive value, defined as the end of the inflation compactification. }

In the second part of the thesis we extend the  DST to the  {Diffusive Action}, that breaks the conservation of the stress energy tensor $T^{\mu\nu}$, and introduces a current source $j^\mu$ \cite{Calogero:2011re,Haba:2009by}. However, the Einstein tensor (and the corresponding matter fields) are covarintly conserved.

 {Reference} \cite{Benisty:2017eqh} extends the DST theory into a Diffusive Action which allows the dark energy and dark matter to exchange energies. This formulates an action principle to the diffusion interaction that was introduced in \cite{Calogero:2011re}  {by hand}. From the asymptotic solution \cite{Benisty:2017eqh}  {one finds} that for some values of the diffusion constant, unlike the standard $\Lambda$CDM model (which has a constant dark energy density and the dark matter decreases like $a^{-3}$) dark energy can slowly decrease, instead of being constant, and dark matter also decreases, but not as fast as $a^{-3}$. This behavior explains the coincidence problem.

While \cite{Benisty:2017eqh} considers one type of the stress energy momentum tensor $T^{\mu\nu}_{(\chi)}$, the last paper in this thesis, \cite{Benisty:2018oyy}, analyzes the complete combination of the stress energy momentum tensor with regards to diffusive interactions.  {Reference} \cite{Benisty:2018oyy} shows analytically and numerically that all of the solutions  {yield the stable $\Lambda$CDM model} with new families of solutions. The general consideration of $T^{\mu\nu}_{(\chi)}$ includes the constants $\lambda_1$ and $\lambda_2$.  {The choice $\lambda_2 = 0 $ corresponds to  {the original papers in the introduction of this thesis} \cite{Benisty:2018qed,Anagnostopoulos:2019myt,Benisty:2018gzx}.} Asymptotically the solutions approach $\Lambda$CDM model and the  {Dynamical Time} approaches the finite value $1/3H_0$. In addition, the case $\lambda_2 =0 $ corresponds to the behavior of non-Lagrangian formulations of dark energy and dark matter that  {suggested in} \cite{Calogero:2011re}.

 {In summery}, the scalar field model of dark energy has a successful description of our  {Universe} without  {a dark matter} component. However under the DST an elegant unification between dark energy and dark matter components emerges naturally. The model is proven to be stable and fits with the data of the late time expansion. Compacitifation with higher dimensions yields inflationary scenarios.  {By breaking} the conservation of the energy momentum tensor $T^{\mu\nu}_{(\chi)}$ diffusion effects are obtained. For  {these} reasons, it seems that DST reveals the capabilities of the scenario and makes it a good candidate for the description of  {Nature}.
\newpage													
\mbox{} 
\newpage

\let\Section\section 
\def\section*#1{\Section{#1}} 

\newpage
\bibliographystyle{unsrt}
\bibliography{ref}

\begin{thebibliography}{10}

\bibitem{Perlmutter:1998np}
S.~Perlmutter et~al.
\newblock {Measurements of $\Omega$ and $\Lambda$ from 42 high redshift
  supernovae}.
\newblock {\em Astrophys. J.}, 517:565--586, 1999.

\bibitem{Riess:1998cb}
Adam~G. Riess et~al.
\newblock {Observational evidence from supernovae for an accelerating universe
  and a cosmological constant}.
\newblock {\em Astron. J.}, 116:1009--1038, 1998.

\bibitem{Angus:2018tko}
C.~R. Angus et~al.
\newblock {Superluminous supernovae from the Dark Energy Survey}.
\newblock {\em Mon. Not. Roy. Astron. Soc.}, 487(2):2215--2241, 2019.

\bibitem{Zhang:2018gbq}
Y.~Zhang et~al.
\newblock {Dark Energy Survey Year 1 results: Detection of Intra-cluster Light
  at Redshift $\sim$ 0.25}.
\newblock {\em Astrophys. J.}, 874(2):165, 2019.

\bibitem{Lambda-CDM-1}
Joshua Frieman, Michael Turner, and Dragan Huterer.
\newblock {Dark Energy and the Accelerating Universe}.
\newblock {\em Ann. Rev. Astron. Astrophys.}, 46:385--432, 2008.

\bibitem{Sandage:2006cv}
A.~Sandage, G.~A. Tammann, A.~Saha, B.~Reindl, F.~D. Macchetto, and N.~Panagia.
\newblock {The Hubble Constant: A Summary of the HST Program for the Luminosity
  Calibration of Type Ia Supernovae by Means of Cepheids}.
\newblock {\em Astrophys. J.}, 653:843--860, 2006.

\bibitem{Riess:2019cxk}
Adam~G. Riess, Stefano Casertano, Wenlong Yuan, Lucas~M. Macri, and Dan
  Scolnic.
\newblock {Large Magellanic Cloud Cepheid Standards Provide a 1\% Foundation
  for the Determination of the Hubble Constant and Stronger Evidence for
  Physics beyond $\Lambda$CDM}.
\newblock {\em Astrophys. J.}, 876(1):85, 2019.

\bibitem{RevModPhys.61.1}
Steven Weinberg.
\newblock The cosmological constant problem.
\newblock {\em Rev. Mod. Phys.}, 61:1--23, Jan 1989.

\bibitem{Verde:2019ivm}
L.~Verde, T.~Treu, and A.~G. Riess.
\newblock {Tensions between the Early and the Late Universe}.
\newblock {\em Nature Astron.}, 3:891, 7 2019.

\bibitem{Weinberg:1988cp}
Steven Weinberg.
\newblock {The Cosmological Constant Problem}.
\newblock {\em Rev. Mod. Phys.}, 61:1--23, 1989.

\bibitem{Martin:2012bt}
Jerome Martin.
\newblock {Everything You Always Wanted To Know About The Cosmological Constant
  Problem (But Were Afraid To Ask)}.
\newblock {\em Comptes Rendus Physique}, 13:566--665, 2012.

\bibitem{Padilla:2015aaa}
Antonio Padilla.
\newblock {Lectures on the Cosmological Constant Problem}.
\newblock 2 2015.

\bibitem{Benisty:2019fzt}
David Benisty, Eduardo~I. Guendelman, and Ofer Lahav.
\newblock {Milky Way and Andromeda past-encounters in different gravity models:
  the impact on the estimated Local Group mass}.
\newblock 4 2019.

\bibitem{Lombriser:2019jia}
Lucas Lombriser.
\newblock {On the cosmological constant problem}.
\newblock {\em Phys. Lett. B}, 797:134804, 2019.

\bibitem{Velten:2014nra}
H.~E.~S. Velten, R.~F. vom Marttens, and W.~Zimdahl.
\newblock {Aspects of the cosmological \textquotedblleft{}coincidence
  problem\textquotedblright{}}.
\newblock {\em Eur. Phys. J. C}, 74(11):3160, 2014.

\bibitem{Dam:2019prv}
Lawrence Dam, Krzysztof Bolejko, and Geraint~F. Lewis.
\newblock {Probing the independence within the dark sector in the fluid
  approximation}.
\newblock {\em JCAP}, 12:030, 2019.

\bibitem{Kisslinger:2019ysx}
Leonard~S. Kisslinger and Debasish Das.
\newblock {A brief review of dark matter}.
\newblock {\em Int. J. Mod. Phys. A}, 34(27):1930013, 2019.

\bibitem{Kamada:2019jch}
Ayuki Kamada, Masaki Yamada, and Tsutomu~T. Yanagida.
\newblock {Unification for darkly charged dark matter}.
\newblock {\em Phys. Rev. D}, 102(1):015012, 2020.

\bibitem{Sola:2016zeg}
Joan Sol\`a.
\newblock {Cosmological constant vis-a-vis dynamical vacuum: bold challenging
  the $\Lambda$CDM}.
\newblock {\em Int. J. Mod. Phys. A}, 31(23):1630035, 2016.

\bibitem{Sola:2016jky}
Joan Sol\`a, Adria G\'omez-Valent, and Javier de~Cruz~P\'erez.
\newblock {First evidence of running cosmic vacuum: challenging the concordance
  model}.
\newblock {\em Astrophys. J.}, 836(1):43, 2017.

\bibitem{Saridakis:2019qwt}
Emmanuel~N. Saridakis, Shynaray Myrzakul, Kairat Myrzakulov, and Koblandy
  Yerzhanov.
\newblock {Cosmological applications of $F(R,T)$ gravity with dynamical
  curvature and torsion}.
\newblock {\em Phys. Rev. D}, 102(2):023525, 2020.

\bibitem{DiValentino:2020srs}
Eleonora Di~Valentino et~al.
\newblock {Cosmology Intertwined IV: The Age of the Universe and its
  Curvature}.
\newblock 9 2020.

\bibitem{DiValentino:2020vvd}
Eleonora Di~Valentino et~al.
\newblock {Cosmology Intertwined III: $f \sigma_8$ and $S_8$}.
\newblock 9 2020.

\bibitem{DiValentino:2020zio}
Eleonora Di~Valentino et~al.
\newblock {Cosmology Intertwined II: The Hubble Constant Tension}.
\newblock 8 2020.

\bibitem{DiValentino:2020vhf}
Eleonora Di~Valentino et~al.
\newblock {Cosmology Intertwined I: Perspectives for the Next Decade}.
\newblock 8 2020.

\bibitem{Abadi:2020hbr}
Tal Abadi and Ely~D. Kovetz.
\newblock {Can conformally coupled modified gravity solve the Hubble tension?}
\newblock {\em Phys. Rev. D}, 103(2):023530, 2021.

\bibitem{DiValentino:2019jae}
Eleonora Di~Valentino, Alessandro Melchiorri, Olga Mena, and Sunny Vagnozzi.
\newblock {Nonminimal dark sector physics and cosmological tensions}.
\newblock {\em Phys. Rev. D}, 101(6):063502, 2020.

\bibitem{Arevalo:2016epc}
Fabiola Arevalo, Antonella Cid, and Jorge Moya.
\newblock {AIC and BIC for cosmological interacting scenarios}.
\newblock {\em Eur. Phys. J. C}, 77(8):565, 2017.

\bibitem{Anagnostopoulos:2017iao}
Fotios~K. Anagnostopoulos and Spyros Basilakos.
\newblock {Constraining the dark energy models with $H(z)$ data: An approach
  independent of $H_0$}.
\newblock {\em Phys. Rev. D}, 97(6):063503, 2018.

\bibitem{Saridakis:2018unr}
Emmanuel~N. Saridakis, Kazuharu Bamba, R.~Myrzakulov, and Fotios~K.
  Anagnostopoulos.
\newblock {Holographic dark energy through Tsallis entropy}.
\newblock {\em JCAP}, 12:012, 2018.

\bibitem{Scherrer:2004au}
Robert~J. Scherrer.
\newblock {Purely kinetic k-essence as unified dark matter}.
\newblock {\em Phys. Rev. Lett.}, 93:011301, 2004.

\bibitem{Arbey:2006it}
Alexandre Arbey.
\newblock {Dark fluid: A Complex scalar field to unify dark energy and dark
  matter}.
\newblock {\em Phys. Rev. D}, 74:043516, 2006.

\bibitem{Chen:2008ft}
Xi-ming Chen, Yun-gui Gong, and Emmanuel~N. Saridakis.
\newblock {Phase-space analysis of interacting phantom cosmology}.
\newblock {\em JCAP}, 04:001, 2009.

\bibitem{Leon:2013qh}
Genly Leon, Joel Saavedra, and Emmanuel~N. Saridakis.
\newblock {Cosmological behavior in extended nonlinear massive gravity}.
\newblock {\em Class. Quant. Grav.}, 30:135001, 2013.

\bibitem{Leon:2012mt}
Genly Leon and Emmanuel~N. Saridakis.
\newblock {Dynamical analysis of generalized Galileon cosmology}.
\newblock {\em JCAP}, 03:025, 2013.

\bibitem{Cai:2015emx}
Yi-Fu Cai, Salvatore Capozziello, Mariafelicia De~Laurentis, and Emmanuel~N.
  Saridakis.
\newblock {f(T) teleparallel gravity and cosmology}.
\newblock {\em Rept. Prog. Phys.}, 79(10):106901, 2016.

\bibitem{Kofinas:2014aka}
Georgios Kofinas, Genly Leon, and Emmanuel~N. Saridakis.
\newblock {Dynamical behavior in $f(T,T_G)$ cosmology}.
\newblock {\em Class. Quant. Grav.}, 31:175011, 2014.

\bibitem{Skugoreva:2014ena}
Maria~A. Skugoreva, Emmanuel~N. Saridakis, and Alexey~V. Toporensky.
\newblock {Dynamical features of scalar-torsion theories}.
\newblock {\em Phys. Rev. D}, 91:044023, 2015.

\bibitem{Bahamonde:2015zma}
Sebastian Bahamonde, Christian~G. B\"ohmer, and Matthew Wright.
\newblock {Modified teleparallel theories of gravity}.
\newblock {\em Phys. Rev. D}, 92(10):104042, 2015.

\bibitem{Bahamonde:2015hza}
Sebastian Bahamonde and Matthew Wright.
\newblock {Teleparallel quintessence with a nonminimal coupling to a boundary
  term}.
\newblock {\em Phys. Rev. D}, 92(8):084034, 2015.
\newblock [Erratum: Phys.Rev.D 93, 109901 (2016)].

\bibitem{Khalifeh:2019zfi}
Ali~Rida Khalifeh, Nicola Bellomo, Jos\'e~Luis Bernal, and Raul Jimenez.
\newblock {Can Dark Matter be Geometry? A Case Study with Mimetic Dark Matter}.
\newblock {\em Phys. Dark Univ.}, 30:100646, 2020.

\bibitem{Solomon:2019qgf}
Adam~R. Solomon, Valeri Vardanyan, and Yashar Akrami.
\newblock {Massive mimetic cosmology}.
\newblock {\em Phys. Lett. B}, 794:135--142, 2019.

\bibitem{Ganz:2018vzg}
Alexander Ganz, Nicola Bartolo, Purnendu Karmakar, and Sabino Matarrese.
\newblock {Gravity in mimetic scalar-tensor theories after GW170817}.
\newblock {\em JCAP}, 01:056, 2019.

\bibitem{Gorji:2018okn}
Mohammad~Ali Gorji, Shinji Mukohyama, Hassan Firouzjahi, and Seyed~Ali
  Hosseini~Mansoori.
\newblock {Gauge Field Mimetic Cosmology}.
\newblock {\em JCAP}, 08:047, 2018.

\bibitem{Arroja:2017msd}
Frederico Arroja, Teppei Okumura, Nicola Bartolo, Purnendu Karmakar, and Sabino
  Matarrese.
\newblock {Large-scale structure in mimetic Horndeski gravity}.
\newblock {\em JCAP}, 05:050, 2018.

\bibitem{GBM:2017lvd}
B.~P. Abbott et~al.
\newblock {Multi-messenger Observations of a Binary Neutron Star Merger}.
\newblock {\em Astrophys. J. Lett.}, 848(2):L12, 2017.

\bibitem{Lombriser:2016yzn}
Lucas Lombriser and Nelson~A. Lima.
\newblock {Challenges to Self-Acceleration in Modified Gravity from
  Gravitational Waves and Large-Scale Structure}.
\newblock {\em Phys. Lett. B}, 765:382--385, 2017.

\bibitem{Creminelli:2017sry}
Paolo Creminelli and Filippo Vernizzi.
\newblock {Dark Energy after GW170817 and GRB170817A}.
\newblock {\em Phys. Rev. Lett.}, 119(25):251302, 2017.

\bibitem{Ezquiaga:2017ekz}
Jose~Mar\'\i{}a Ezquiaga and Miguel Zumalac\'arregui.
\newblock {Dark Energy After GW170817: Dead Ends and the Road Ahead}.
\newblock {\em Phys. Rev. Lett.}, 119(25):251304, 2017.

\bibitem{Ratra:1987rm}
Bharat Ratra and P.~J.~E. Peebles.
\newblock {Cosmological Consequences of a Rolling Homogeneous Scalar Field}.
\newblock {\em Phys. Rev. D}, 37:3406, 1988.

\bibitem{Caldwell:1997ii}
R.~R. Caldwell, Rahul Dave, and Paul~J. Steinhardt.
\newblock {Cosmological imprint of an energy component with general equation of
  state}.
\newblock {\em Phys. Rev. Lett.}, 80:1582--1585, 1998.

\bibitem{Zlatev:1998tr}
Ivaylo Zlatev, Li-Min Wang, and Paul~J. Steinhardt.
\newblock {Quintessence, cosmic coincidence, and the cosmological constant}.
\newblock {\em Phys. Rev. Lett.}, 82:896--899, 1999.

\bibitem{Chiba:1999ka}
Takeshi Chiba, Takahiro Okabe, and Masahide Yamaguchi.
\newblock {Kinetically driven quintessence}.
\newblock {\em Phys. Rev. D}, 62:023511, 2000.

\bibitem{Barreiro:1999zs}
T.~Barreiro, Edmund~J. Copeland, and N.~J. Nunes.
\newblock {Quintessence arising from exponential potentials}.
\newblock {\em Phys. Rev. D}, 61:127301, 2000.

\bibitem{Bento:2002ps}
M.~C. Bento, O.~Bertolami, and A.~A. Sen.
\newblock {Generalized Chaplygin gas, accelerated expansion and dark energy
  matter unification}.
\newblock {\em Phys. Rev. D}, 66:043507, 2002.

\bibitem{dePutter:2007ny}
Roland de~Putter and Eric~V. Linder.
\newblock {Kinetic k-essence and Quintessence}.
\newblock {\em Astropart. Phys.}, 28:263--272, 2007.

\bibitem{Tsujikawa:2013fta}
Shinji Tsujikawa.
\newblock {Quintessence: A Review}.
\newblock {\em Class. Quant. Grav.}, 30:214003, 2013.

\bibitem{Babichev:2018twg}
Eugeny Babichev, Sabir Ramazanov, and Alexander Vikman.
\newblock {Recovering $P(X)$ from a canonical complex field}.
\newblock {\em JCAP}, 11:023, 2018.

\bibitem{Kehayias:2019gir}
John Kehayias and Robert~J. Scherrer.
\newblock {New generic evolution for $k$ -essence dark energy with $w \approx
  -1$}.
\newblock {\em Phys. Rev. D}, 100(2):023525, 2019.

\bibitem{Oikonomou:2019muq}
V.~K. Oikonomou and N.~Chatzarakis.
\newblock {The Phase Space of $k$-Essence $f(R)$ Gravity Theory}.
\newblock {\em Nucl. Phys. B}, 956:115023, 2020.

\bibitem{Chakraborty:2019swx}
Abhijit Chakraborty, Anandamohan Ghosh, and Narayan Banerjee.
\newblock {Dynamical systems analysis of a k -essence model}.
\newblock {\em Phys. Rev. D}, 99(10):103513, 2019.

\bibitem{Chervon:2019sey}
Sergei Chervon, Igor Fomin, Valerian Yurov, and Artyom Yurov.
\newblock {\em {Scalar Field Cosmology}}, volume~13 of {\em Series on the
  Foundations of Natural Science and Technology}.
\newblock WSP, Singapur, 2019.

\bibitem{Gao:2009me}
Changjun Gao, Martin Kunz, Andrew~R. Liddle, and David Parkinson.
\newblock {Unified dark energy and dark matter from a scalar field different
  from quintessence}.
\newblock {\em Phys. Rev. D}, 81:043520, 2010.

\bibitem{Guendelman:2009ck}
E.~I. Guendelman.
\newblock {Gravitational Theory with a Dynamical Time}.
\newblock {\em Int. J. Mod. Phys. A}, 25:4081--4099, 2010.

\bibitem{Friedman:1922kd}
A.~Friedman.
\newblock {On the Curvature of space}.
\newblock {\em Z. Phys.}, 10:377--386, 1922.

\bibitem{Aghanim:2018eyx}
N.~Aghanim et~al.
\newblock {Planck 2018 results. VI. Cosmological parameters}.
\newblock {\em Astron. Astrophys.}, 641:A6, 2020.

\bibitem{Guth:1980zm}
Alan~H. Guth.
\newblock {The Inflationary Universe: A Possible Solution to the Horizon and
  Flatness Problems}.
\newblock {\em Phys. Rev. D}, 23:347--356, 1981.

\bibitem{Guth:1982ec}
Alan~H. Guth and S.~Y. Pi.
\newblock {Fluctuations in the New Inflationary Universe}.
\newblock {\em Phys. Rev. Lett.}, 49:1110--1113, 1982.

\bibitem{Starobinsky:1979ty}
Alexei~A. Starobinsky.
\newblock {Spectrum of relict gravitational radiation and the early state of
  the universe}.
\newblock {\em JETP Lett.}, 30:682--685, 1979.

\bibitem{Kazanas:1980tx}
D.~Kazanas.
\newblock {Dynamics of the Universe and Spontaneous Symmetry Breaking}.
\newblock {\em Astrophys. J. Lett.}, 241:L59--L63, 1980.

\bibitem{Starobinsky:1980te}
Alexei~A. Starobinsky.
\newblock {A New Type of Isotropic Cosmological Models Without Singularity}.
\newblock {\em Phys. Lett. B}, 91:99--102, 1980.

\bibitem{Linde:1981mu}
Andrei~D. Linde.
\newblock {A New Inflationary Universe Scenario: A Possible Solution of the
  Horizon, Flatness, Homogeneity, Isotropy and Primordial Monopole Problems}.
\newblock {\em Phys. Lett. B}, 108:389--393, 1982.

\bibitem{Albrecht:1982wi}
Andreas Albrecht and Paul~J. Steinhardt.
\newblock {Cosmology for Grand Unified Theories with Radiatively Induced
  Symmetry Breaking}.
\newblock {\em Phys. Rev. Lett.}, 48:1220--1223, 1982.

\bibitem{Barrow:1983rx}
John~D. Barrow and A.~C. Ottewill.
\newblock {The Stability of General Relativistic Cosmological Theory}.
\newblock {\em J. Phys. A}, 16:2757, 1983.

\bibitem{Blau:1986cw}
Steven~K. Blau, E.~I. Guendelman, and Alan~H. Guth.
\newblock {The Dynamics of False Vacuum Bubbles}.
\newblock {\em Phys. Rev. D}, 35:1747, 1987.

\bibitem{Copeland:2006wr}
Edmund~J. Copeland, M.~Sami, and Shinji Tsujikawa.
\newblock {Dynamics of dark energy}.
\newblock {\em Int. J. Mod. Phys. D}, 15:1753--1936, 2006.

\bibitem{Frieman:2008sn}
Joshua Frieman, Michael Turner, and Dragan Huterer.
\newblock {Dark Energy and the Accelerating Universe}.
\newblock {\em Ann. Rev. Astron. Astrophys.}, 46:385--432, 2008.

\bibitem{Abbott:2017wau}
T.~M.~C. Abbott et~al.
\newblock {Dark Energy Survey year 1 results: Cosmological constraints from
  galaxy clustering and weak lensing}.
\newblock {\em Phys. Rev. D}, 98(4):043526, 2018.

\bibitem{Abbott:2018xao}
T.~M.~C. Abbott et~al.
\newblock {Dark Energy Survey Year 1 Results: Constraints on Extended
  Cosmological Models from Galaxy Clustering and Weak Lensing}.
\newblock {\em Phys. Rev. D}, 99(12):123505, 2019.

\bibitem{Huang:2020tpm}
Hung-Jin Huang et~al.
\newblock {Dark Energy Survey Year 1 Results: Constraining Baryonic Physics in
  the Universe}.
\newblock {\em Mon. Not. Roy. Astron. Soc.}, 502(4):6010--6031, 2021.

\bibitem{Peebles:1987ek}
P.~J.~E. Peebles and Bharat Ratra.
\newblock {Cosmology with a Time Variable Cosmological Constant}.
\newblock {\em Astrophys. J. Lett.}, 325:L17, 1988.

\bibitem{Wetterich:1994bg}
Christof Wetterich.
\newblock {The Cosmon model for an asymptotically vanishing time dependent
  cosmological 'constant'}.
\newblock {\em Astron. Astrophys.}, 301:321--328, 1995.

\bibitem{Frieman:1995pm}
Joshua~A. Frieman, Christopher~T. Hill, Albert Stebbins, and Ioav Waga.
\newblock {Cosmology with ultralight pseudo Nambu-Goldstone bosons}.
\newblock {\em Phys. Rev. Lett.}, 75:2077--2080, 1995.

\bibitem{Ferreira:1997au}
Pedro~G. Ferreira and Michael Joyce.
\newblock {Structure formation with a selftuning scalar field}.
\newblock {\em Phys. Rev. Lett.}, 79:4740--4743, 1997.

\bibitem{Viana:1997mt}
Pedro T.~P. Viana and Andrew~R. Liddle.
\newblock {Perturbation evolution in cosmologies with a decaying cosmological
  constant}.
\newblock {\em Phys. Rev. D}, 57:674--684, 1998.

\bibitem{Copeland:1997et}
Edmund~J. Copeland, Andrew~R Liddle, and David Wands.
\newblock {Exponential potentials and cosmological scaling solutions}.
\newblock {\em Phys. Rev. D}, 57:4686--4690, 1998.

\bibitem{Martin:2008qp}
Jerome Martin.
\newblock {Quintessence: a mini-review}.
\newblock {\em Mod. Phys. Lett. A}, 23:1252--1265, 2008.

\bibitem{Benisty:2016ybt}
David Benisty and E.~I. Guendelman.
\newblock {Radiation Like Scalar Field and Gauge Fields in Cosmology for a
  theory with Dynamical Time}.
\newblock {\em Mod. Phys. Lett. A}, 31(33):1650188, 2016.

\bibitem{Calogero:2011re}
Simone Calogero.
\newblock {A kinetic theory of diffusion in general relativity with
  cosmological scalar field}.
\newblock {\em JCAP}, 11:016, 2011.

\bibitem{Haba:2009by}
Z.~Haba.
\newblock {Relativistic diffusion with friction on a pseudoriemannian
  manifold}.
\newblock {\em Class. Quant. Grav.}, 27:095021, 2010.

\bibitem{Benisty:2018qed}
David Benisty and Eduardo~I. Guendelman.
\newblock {Unified dark energy and dark matter from dynamical spacetime}.
\newblock {\em Phys. Rev. D}, 98(2):023506, 2018.

\bibitem{Anagnostopoulos:2019myt}
Fotios~K. Anagnostopoulos, David Benisty, Spyros Basilakos, and Eduardo~I.
  Guendelman.
\newblock {Dark energy and dark matter unification from dynamical space time:
  observational constraints and cosmological implications}.
\newblock {\em JCAP}, 06:003, 2019.

\bibitem{Benisty:2018gzx}
David Benisty and Eduardo~I. Guendelman.
\newblock {Inflation compactification from dynamical spacetime}.
\newblock {\em Phys. Rev. D}, 98(4):043522, 2018.

\bibitem{Benisty:2017eqh}
David Benisty and E.~I. Guendelman.
\newblock {Interacting Diffusive Unified Dark Energy and Dark Matter from
  Scalar Fields}.
\newblock {\em Eur. Phys. J. C}, 77(6):396, 2017.

\bibitem{Benisty:2018oyy}
David Benisty, Eduardo Guendelman, and Zbigniew Haba.
\newblock {Unification of dark energy and dark matter from diffusive
  cosmology}.
\newblock {\em Phys. Rev. D}, 99(12):123521, 2019.
\newblock [Erratum: Phys.Rev.D 101, 049901 (2020)].

\bibitem{Aghanim:2019ame}
N.~Aghanim et~al.
\newblock {Planck 2018 results. V. CMB power spectra and likelihoods}.
\newblock {\em Astron. Astrophys.}, 641:A5, 2020.

\bibitem{Freedman:2000cf}
W.~L. Freedman et~al.
\newblock {Final results from the Hubble Space Telescope key project to measure
  the Hubble constant}.
\newblock {\em Astrophys. J.}, 553:47--72, 2001.

\bibitem{Riess:2016jrr}
Adam~G. Riess et~al.
\newblock {A 2.4\% Determination of the Local Value of the Hubble Constant}.
\newblock {\em Astrophys. J.}, 826(1):56, 2016.

\bibitem{Riess:2020fzl}
Adam~G. Riess, Stefano Casertano, Wenlong Yuan, J.~Bradley Bowers, Lucas Macri,
  Joel~C. Zinn, and Dan Scolnic.
\newblock {Cosmic Distances Calibrated to 1\% Precision with Gaia EDR3
  Parallaxes and Hubble Space Telescope Photometry of 75 Milky Way Cepheids
  Confirm Tension with $\Lambda$CDM}.
\newblock {\em Astrophys. J. Lett.}, 908(1):L6, 2021.

\end{thebibliography}

\newpage													
\mbox{} 
\newpage
\includepdf[pages=-]{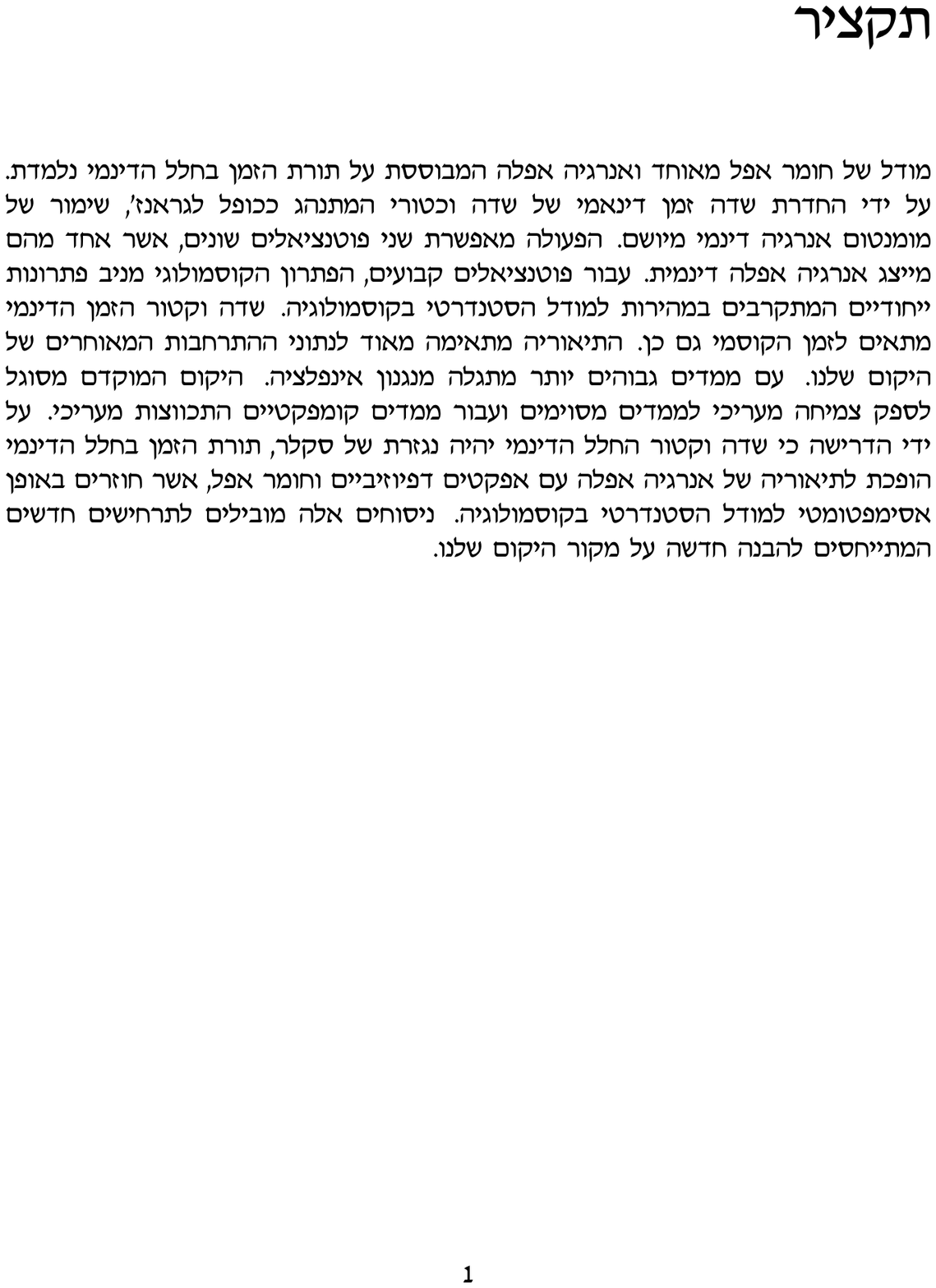}

\end{document}